\documentclass{svmult}
\usepackage{makeidx}
\usepackage{graphicx}
\usepackage{multicol}
\usepackage[bottom]{footmisc}
\makeindex

\begin{document}

\title*{Heterogeneous Economic Networks}
\author{
Wataru Souma\inst{1}\and
Yoshi Fujiwara\inst{2}\and
Hideaki Aoyama\inst{3}}
\institute{
ATR Network Informatics Laboratories, Kyoto 619-0288, Japan. \texttt{souma@atr.jp}\and
ATR Network Informatics Laboratories, Kyoto 619-0288, Japan. \texttt{yfujiwar@atr.jp}\and
Department of Physics, Graduate School of Science, Kyoto University, Yoshida,
Kyoto 606-8501, Japan. \texttt{aoyama@phys.h.kyoto-u.ac.jp}}
\maketitle

\vspace{5mm}
{\bfseries Summary.}
The Japanese shareholding network at the end of March 2002 is studied.
To understand the characteristics of this network intuitively, we 
visualize it as a directed graph and an adjacency matrix.
Especially detailed features of networks concerned with the automobile industry sector
are discussed by using the visualized networks.
The shareholding network is also considered as an undirected graph, because many
quantities characterizing networks are defined for undirected cases.
For this undirected shareholding network, we show that a degree distribution is well fitted by a power law
function with an exponential tail.  The exponent in the power law range is $\gamma=1.8$.
We also show that the spectrum of this network follows asymptotically the power law distribution
with the exponent $\delta=2.6$. 
By comparison with $\gamma$ and $\delta$, we find a scaling relation $\delta=2\gamma-1$.
The reason why this relation holds is attributed to the local tree-like structure of networks.
To clarify this structure, the correlation between degrees and clustering coefficients is considered.
We show that this correlation is negative and fitted by the power law function with the exponent
$\alpha=1.1$. This guarantees the local tree-like structure of the network and suggests
the existence of a hierarchical structure.
We also show that the degree correlation is negative and follows the power law function
with the exponent $\nu=0.8$. 
This indicates a degree-nonassortative network, in which hubs are not directly
connected with each other.
To understand these features of the network from the viewpoint of a company's growth,
we consider the correlation between the degree and the company's total assets and age.
It is clarified that the degree and the company's total assets correlate strongly, but
the degree and the company's age have no correlation.

\vspace{5mm}

\section{Introduction}
\label{sec:1}
The economy is regarded as a set of activities of heterogeneous agents in complex networks.
However, many traditional studies in economics are for the activities of
homogeneous agents in simple networks, where we call regular networks and random networks
simple networks. To overcome such an unrealistic situation many efforts have been made,
and a viewpoint of heterogeneous agents has emerged.
However, simple networks are adapted in many of the studies of heterogeneous agents.
Hence it is important to introduce the true structure of real world networks.
Recently, the study of complex networks has revealed the true structure of
real world networks: WWW, the Internet, social networks, biological networks, etc.
\cite{ab}\cite{barabasi}\cite{dm}\cite{newmanSIMA}.
However the true structure of the economic network is not well known.
Hence the purpose of this study is to reveal it.

As is commonly known, if we intend to discuss networks, we must define
the nodes and edges. Here, edges represent the relationship between nodes.
In business networks, the candidates for the nodes are individuals, companies, industry categories,
countries, etc. In this study we consider companies as nodes.
Hence, in the next step, we must define the relationship between companies.
To define it, we use three viewpoints: ownership, governance, and activity.
The ownership is characterized by the shareholding of companies, and the
networks constructed by this relationship are considered in this article.
The governance is characterized by the interlocking of directors,
and networks of this type are frequently
represented by a bipartite graph that is constructed with corporate boards and directors.
The activity networks are characterized by many relationships: trade, collaboration, etc.

Although we use tree point of view, these have relations with each other.
For example, if the owners of a company change, then the directors of that company will change.
If the directors of the company change, then the activities of the company will change.
If the activities of the company change, then the decisions of the owners and directors will change,
and sometimes the owners and the directors will change.

In this article we consider Japanese shareholding networks
at the end of March 2002 (see Ref.~\cite{gbcsc} for shareholding
networks in MIB, NYSE, and NASDAQ).
In this study we use data which is published by TOYO KEIZAI INC.
This data provides lists of shareholders for 2,765 companies that are listed on
the stock market or the over-the-counter market. 
Almost all of the shareholders are non-listed financial institutions (commercial banks, trust banks,
and insurance companies) and listed non-financial companies. 
In this article we ignore shares held by officers and other individuals.
The lengths of the shareholder lists vary with the companies.
The most comprehensive lists contain information on the top 30 shareholders.
Based on this data we construct a shareholding network.
 
This paper is organized as follows. In Sec.~\ref{sec:2}, we consider the visualization of
the shareholding network as a directed graph and an adjacency matrix.
The visualization of networks is a primitive, but powerful tool to intuitively understand the
characteristics of networks.
As an example, we especially consider networks concerned with the automobile industry sector.
In Sec.~\ref{sec:3}, we treat the shareholding network as an undirected network.
This is because many useful quantities characterizing networks are defined
for undirected cases.
We consider the degree distribution, spectrum, degree correlation, and
the correlation between the degree and the clustering coefficient.
In Sec.~\ref{sec:4}, we discuss correlations between the degree and the company's total assets
and age.
Section \ref{sec:5} is devoted to a summary and discussion.

\section{Directed networks}
\label{sec:2}
\begin{figure}[!t]
\centerline{
\begin{minipage}{0.43\linewidth}
\includegraphics[width=\linewidth]{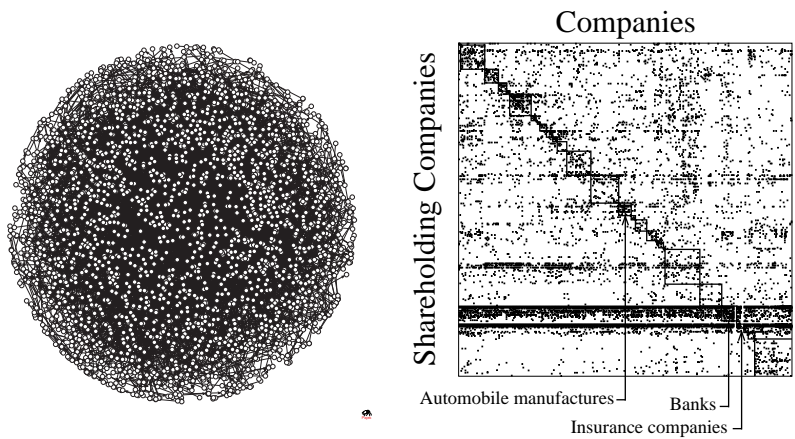}
\end{minipage}
\hspace{3mm}
\begin{minipage}{0.43\linewidth}
\includegraphics[width=\linewidth]{adj_bound_v2.eps}
\end{minipage}
}
\caption{A visualization of the Japanese shareholding network at the end of March 2002 (left)
and a corresponding adjacency matrix (right).}
\label{fig:1} 
\end{figure}

If we draw edges from shareholders to companies, we can obtain
shareholding networks as directed graphs.
The primitive way to study networks is to use a visualization of them.
The visualization of the Japanese shareholding network at the end of March 2002
is shown in the left panel of Fig.~\ref{fig:1}.
This figure is drawn by Pajek, which is a program for analyzing large
networks \cite{bm}.
In this figure, the open circles correspond to companies, and arrows are drawn
based on shareholding relationships.
The network is constructed from 3,152 nodes and 23,064 arrows.
This figure is beautiful, but it is difficult to obtain characteristics of this network.

Frequently, networks are represented by adjacency matrices.
Here the adjacency matrix for the directed graph,  $M^d_{ij}$, is chosen based on the shareholding
relation: If the $i$-th company is a shareholder of the $j$-th company,
we assume $M^d_{ij}=1$; otherwise, $M^d_{ij}=0$.
The size of this matrix (network) is $N=3,152$.
The adjacency matrix is shown in the right panel of Fig.~\ref{fig:1}.
In this figure, the rows correspond to companies and the columns correspond to shareholding companies.
The list of companies and the list of shareholding companies are the same.
In this figure, the black dots correspond to $M^d_{ij}=1$, and the others correspond to $M^d_{ij}=0$.
The number of black dots is 23,063.

To define this adjacency matrix we arranged the order of companies
according to the company's code number that is defined based on industry categories.
The solid lines in this figure indicate industry categories. 
We make two observations:
(i) Dots are distributed in all industry categories in the ranges
where financial institutions (banks and insurance companies) are the shareholders;
(ii) The density of the black dots is relatively high in each box, except for
the financial sector.
This indicates that we frequently find companies and shareholders in the same
industry category for non-financial firms.
Hence this network shows "assortative mixing" on the industry category, except for the
financial institutions. The concept of (dis)assortativity is explained in
Ref.~\cite{newman}.

\subsection{Shareholding network constructed from companies in the  automobile industry sector}
\label{sec:21}

\begin{figure}[!t]
\centerline{
\includegraphics[width=0.7\linewidth]{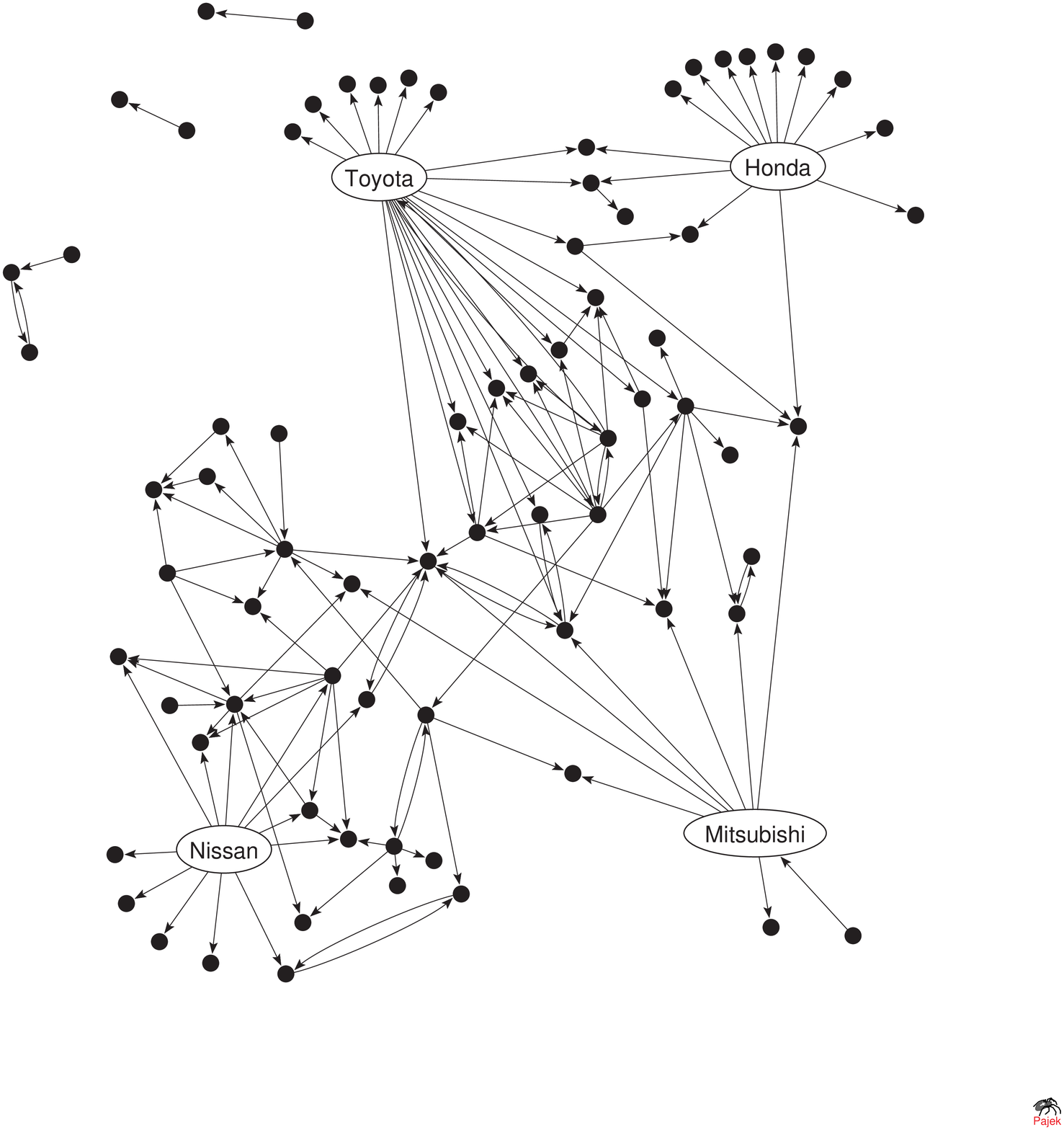}
}
\caption{A shareholding network constructed from companies in
the automobile industry sector.}
\label{fig:2}
\end{figure}

Here, we consider a network constructed
from companies belonging to the automobile industry sector.
This sector corresponds to the range shown in the right panel in Fig.~\ref{fig:1}.
The visualization of this network is shown in Fig.~\ref{fig:2}.
In this figure, we include company names only for four major automobile industry companies:
Toyota Motor, Nissan Motor, Honda Motor, and Mitsubishi Motor.

If the direction of the arrows is ignored, then Toyota, Honda, and Mitsubishi are connected to each other
with two edges through the shortest path.
Nissan and Toyota are connected with three edges through the shortest path,
and this is also applicable to the case of Nissan and Mitsubishi.
However, we need five edges to connect Nissan and Honda through the shortest path.
In addition, this path must run through Toyota or Mitsubishi.
We presently have no idea how to explain such a network structure,
but we believe that a time series analysis of networks is important.

This figure shows that these four major companies have many edges.
In the graph theory, the degree of a node is defined by the number of other nodes
to which it is attached.
The distribution of the degree is an important quantity to study networks \cite{ba}.
It is well known that the degree distribution of a regular network is represented by 
the $\delta$-function, because each node has the same degree in the regular network.
It is also well known that the degree distribution of random networks,
which are proposed by Erd\"{o}s and R\'{e}ny, follows the Poisson distribution.
A node with a large degree is called a hub. Hence these four major automobile industry
companies are hubs in the network.
In Sec.~\ref{sec:31}, the details of a study about the degree distribution of the network are
explained.

In this figure, we can find that almost no hubs are directly connected.
These hubs are mediated by companies, which have a small degree.
Networks with such a characteristic are called uncorrelated networks
or degree-nonassortative networks \cite{newman},
and are characterized by a negative correlation between degrees.
This is explained in terms of a degree correlation \cite{pvv} in Sec.~\ref{sec:32}. 
Intuitively, this nature of the shareholding network is different from that of human networks.
This is because, in human networks, for example in the friendship network,
the hubs in the network correspond to persons with many friends,
and are also friends with each other with a high probability.
Networks with such characteristics are called correlated networks or degree-assortative
networks, and are characterized by a positive correlation between degrees.

Suppose that a node has $k$ neighbors; then at most $k(k-1)/2$ edges
can exist to connect the neighbors with each other. Hence the possible number of
triangles (minimum loops) containing
this node is also $k(k-1)/2$.
The ratio between the possible number
of triangles and that of the actually existing triangles defines a clustering coefficient \cite{ws}. 
As we can see in this figure, the clustering coefficient is small for hubs,
while it is large for nodes with a small degree.
Hence it is expected that degrees and clustering coefficients
are negatively correlated.
The details are explained in Sec.~\ref{sec:33}.

This figure also shows that the network is constructed from subgraphs:
triangles, squares, pentagons, etc.
However, not all subgraphs occur with equal frequency. Hence if networks contain
some subgraphs as compared to randomized networks, these subgraphs characterize the networks.
These characteristic subgraphs are the building blocks of networks, and are
called network motifs \cite{msikca}\cite{smma}.
Although network motifs are not discussed in this article, the spectrum of the network
is considered. This is a primitive way to study subgraphs, and the details are
discussed in Sec.~\ref{sec:34}.

\subsection{Shareholding network constructed from edges drawn from the outside of the automobile industry
sector to the inside of it}
\label{sec:22}
\begin{figure}[!t]
\centerline{
\includegraphics[width=0.7\linewidth]{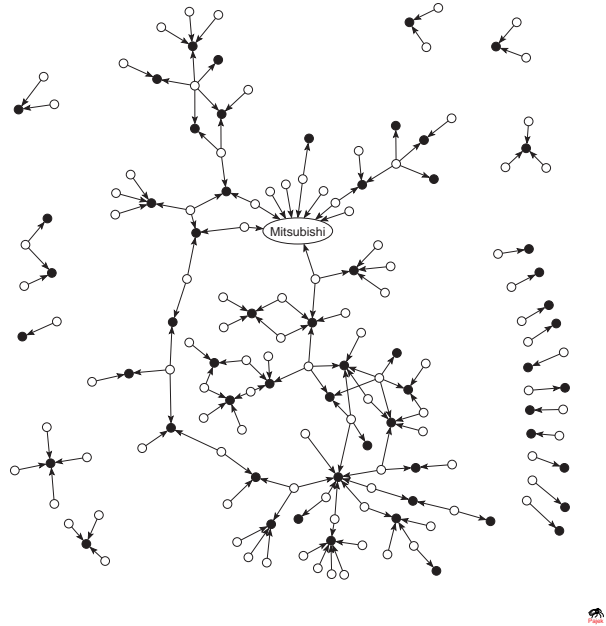}
}
\caption{A shareholding network constructed from edges drawn from the outside of the automobile industry
sector to the inside of it. The open circles correspond to non-automobile industry companies,
and the filled circles correspond to automobile industry companies. Arrows are drawn from the open
circles to the filled circles}
\label{fig:3}
\end{figure}
Figure \ref{fig:3} is constructed from edges drawn from the outside of
the automobile industry sector to the inside of it.
To draw this figure, for simplicity, we ignored the arrows connecting financial institutions
and automobile industry companies.
In this figure the open circles correspond to non-automobile industry companies,
and the filled circles correspond to automobile industry companies.
Arrows are drawn from the open circles to the filled circles.
The network is divided into many subgraphs, but there exists one large lump.
This figure contains only Mitsubishi Motors of the four major automobile
companies. This means that Mitsubishi Motor
is governed by companies outside the automobile industry sector.  

\subsection{Shareholding network constructed from edges drawn from the inside of the automobile industry
sector to the outside of it}
\label{sec:23}
\begin{figure}[!t]
\centerline{
\includegraphics[width=0.8\linewidth]{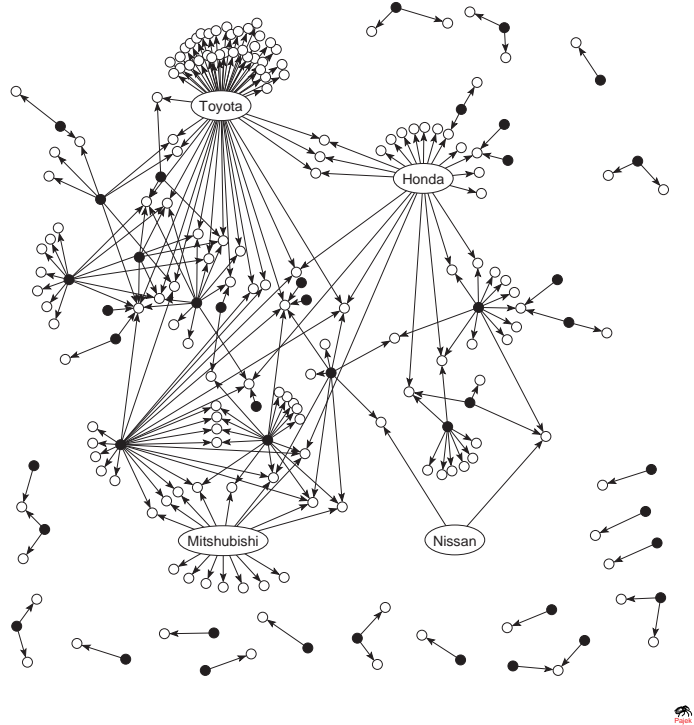}
}
\caption{A shareholding network constructed from edges drawn from the inside of the automobile industry
sector to the outside of it. The open circles correspond to non-automobile industry companies,
and the filled circles correspond to automobile industry companies. Arrows are drawn from the filled
circles to the open circles.}
\label{fig:4}
\end{figure}

Figure \ref{fig:4} is constructed from edges drawn from the inside of
the automobile industry sector to the outside of it.
As in the previous case, the open circles correspond to non-automobile industry companies,
and the filled circles correspond to automobile industry companies.
In this case arrows are drawn from the filled circles to the open circles.
The network is divided into many subgraphs, but a large lump exists.
We can see that major automobile industry companies are also major shareholders of
non-automobile industry companies, except for Nissan Motor.
It is expected that such a structure emerged after the year 1999
when Nissan and Renault announced their strategic alliance.
Comparing this figure and  Figs.~\ref{fig:2} and \ref{fig:3} makes us believe that
Toyota, Honda and Nissan are leaders in the automobile industry sector, and especially
Toyota and Honda are also leaders in the Japanese economy.

\section{Undirected network}
\label{sec:3}
As shown in Sec.~\ref{sec:2}, the visualization of networks is a powerful method to understand
the characteristics of networks. However this method is not always useful.
Hence many quantities have been proposed to obtain the characteristics of networks .
In the previous section, the shareholding network is represented by a directed graph, but we consider
it as an undirected network in this section.
This is because many quantities characterizing networks are for undirected cases.
In this case, the adjacency matrix $M^u_{ij}$ is chosen: If the $i$-th company is a shareholder
of the $j$-th company, we assume $M^u_{ij}=M^u_{ji}=1$; otherwise, $M^u_{ij}=0$.
Hence this matrix is symmetrical.

\subsection{Degree distribution}
\label{sec:31}
\begin{figure}[!t]
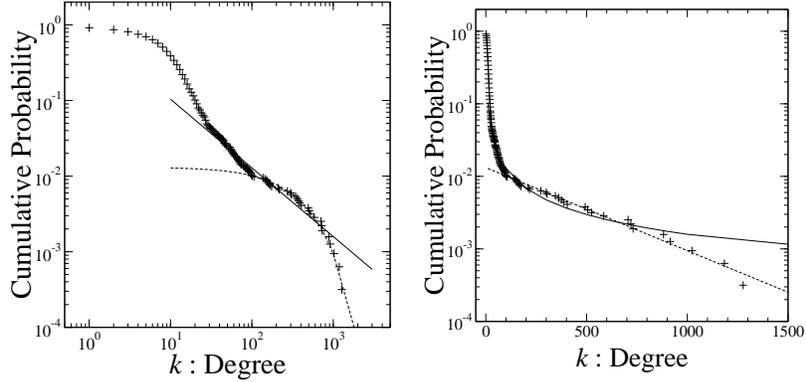

\centerline{
\begin{minipage}{0.43\linewidth}
\includegraphics[width=\linewidth]{tdeg_ll.eps}
\end{minipage}
\hspace{2mm}
\begin{minipage}{0.43\linewidth}
\includegraphics[width=\linewidth]{tdeg_sl.eps}
\end{minipage}
}
\caption{A log-log plot (left) and a semi-log plot (right) of the degree distribution .
In these figures, the solid lines
correspond to the fitting by the power law function $p(k)\approx k^{-\gamma}$ with the exponent
$\gamma=1.8$, and the dashed lines correspond to the fitting by
the exponential function $p(k)\approx\exp\{-\beta(k-k_0)\}$.}
\label{fig:5}
\end{figure}

A degree is the number of edges that attach to a node.
In terms of the adjacency matrix, the degree of node $i$, $k_i$,
is defined by $k_i\equiv\sum_{j=1}^{N}M^u_{ij}.$
The log-log plot of the degree distribution is shown in the left
panel of Fig.~\ref{fig:5}, and the semi-log plot is shown in the right
panel of Fig.~\ref{fig:5}.
In this figure, the horizontal axis corresponds to the degree and
the vertical axis corresponds to the cumulative probability.

In these figures, the solid lines correspond to the linear fitting with the least square method
by the power law function, and the dashed lines correspond to that by the exponential function.
As we can see, the degree distribution follows the power law distribution with
an exponential tail.
The probability density function (PDF) $p(k)$ in the power law range is given by
\[
p(k)\propto k^{-\gamma},
\]
where the exponent $\gamma=1.8$, and that in the exponential range is given by
\[
p(k)\propto e^{-\beta(k-k_0)}.
\]
The exponential range is constructed of financial institutions, and the power law
range is mainly constructed of non-financial firms. It has also been shown that the degree distributions
of different networks in the economy also show power law distributions \cite{souma}.

\subsection{Degree correlation}
\label{sec:32}
To obtain more detailed characteristics of networks, the degree
correlation has been considered \cite{pvv}.
The nearest neighbors' average degree of nodes with degree $k$,
$\langle k_{\rm nn} \rangle$, is
defined by
\[
\langle k_{\rm nn} \rangle=\sum_{k}k'p_c(k'|k),
\]
where $p_c(k'|k)$ is the conditional probability that a link belonging to a
node with degree $k$ points to a node with degree $k'$.

\begin{figure}[!t]
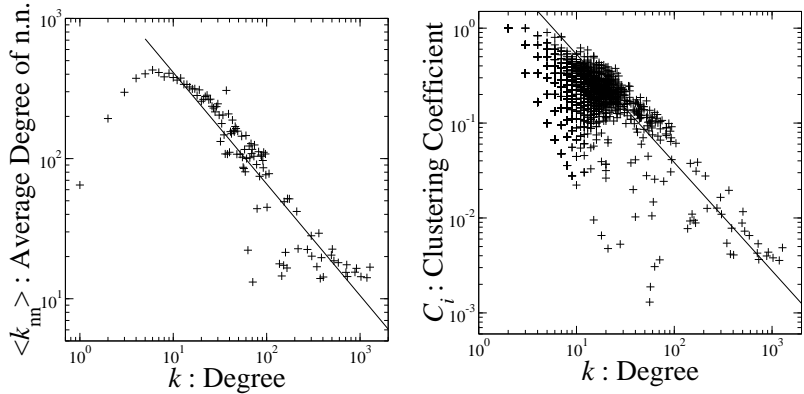

\centerline{
\begin{minipage}{0.43\linewidth}
\includegraphics[width=\linewidth]{deg_corr.eps}
\end{minipage}
\hspace{2mm}
\begin{minipage}{0.43\linewidth}
\includegraphics[width=\linewidth]{deg_cc.eps}
\end{minipage}
}
\caption{A log-log plot of the degree correlation (left) and that of the correlation between
the degree and the clustering coefficient (right).
In the left panel, the solid line corresponds to the fitting by the power law function
$\langle k_{\rm nn}\rangle (k)\approx k^{-\nu}$ with the exponent
$\nu=0.8$. In the right panel, the solid line corresponds to the fitting by the power law function
$C(k)\approx k^{-\alpha}$ with the exponent $\alpha=1.1$.}
\label{fig:6}
\end{figure}

The log-log plot of degree correlation is shown in the left panel of Fig.~\ref{fig:6}.
In this figure, the horizontal axis corresponds to the degree and
the vertical axis corresponds to
$\langle k_{\rm nn} \rangle$, i.e., the nearest neighbors' average degree of nodes with degree $k$.
We find that the high degree range has a small value of $\langle k_{\rm nn} \rangle$.
This means that hubs are not directly connected with each other in this network.
Networks with this characteristic are called uncorrelated networks, which
are also found in biological networks and WWW.
On the other hand, networks with a positive correlation are called correlated networks,
and are found in social and architectural networks.

In Fig.~\ref{fig:6}, the solid line is
the fitting by the power law function in the tail part with the least square method:
\[
\langle k_{\rm nn}\rangle(k) \propto k^{-\nu},
\]
where the exponent $\nu=0.8$.

\subsection{Clustering coefficient}
\label{sec:33}
Cliques in networks are quantified by a clustering coefficient \cite{ws}.
Suppose that a node $i$ has $k_i$ edges; then at most $k_i(k_i-1)/2$ edges
can exist between them. The clustering coefficient of node $i$, $C_i$, is the fraction of
these allowable edges that actually exist $e_i$:
\[
C_i=\frac{2e_i}{k_i(k_i-1)}.
\]
The clustering coefficient is approximately equal to the probability of
finding triangles in the network. The triangle is the minimum loop.
Hence if node $i$ has a small value of $C_i$, then the probability of finding loops
around this node is low. This means that the network around this node is locally tree-like.
The correlation between $k_i$ and $C_i$ is shown in the right panel of Fig.~\ref{fig:6}.
This figure shows that clustering coefficients have a small value 
in the high degree range.
This means that the shareholding network has a local
tree-like structure asymptotically.

The solid line in the right panel of Fig.~\ref{fig:6} is the linear fitting by the power law function
with the least square method:
\[
C(k)\propto k^{-\alpha},
\]
where the exponent $\alpha=1.1$. Such a scaling property of the distribution of
clustering coefficients is also observed in biological networks, and
motivates the concept of hierarchical networks \cite{rsmob}\cite{rb}. 

\subsection{Spectrum}
\label{sec:34}
\begin{figure}[!t]
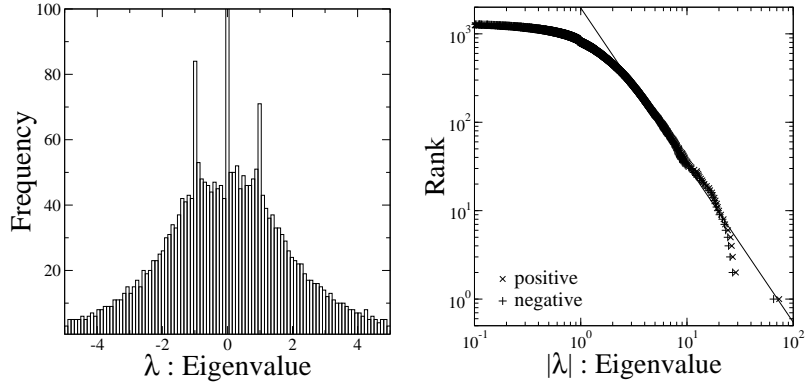

\centerline{
\begin{minipage}{0.43\linewidth}
\includegraphics[width=\linewidth]{eval_origin.eps}
\end{minipage}
\hspace{2mm}
\begin{minipage}{0.43\linewidth}
\includegraphics[width=\linewidth]{eval_tail.eps}
\end{minipage}
}
\caption{A distribution of eigenvalues around the origin (left) and
a log-log plot of the distribution of the absolute value of eigenvalues (right).
In the right panel, the solid line corresponds to the
fitting by the power law function $\rho(\lambda)\approx |\lambda|^{-\delta}$ with the exponent
$\delta=2.6$.}
\label{fig:7}
\end{figure}

Here we discuss the spectrum of the network, i.e.,
the distribution of eigenvalues of the adjacency matrix.
The distribution around the origin is shown in the left
panel of Fig.~\ref{fig:7}. In this figure the horizontal axis is an
eigenvalue $\lambda_i$ and the vertical axis is a frequency.
As is well known, if the network is completely random the distribution is
explained by Wigner's semi-circle law.
However, Fig.~ \ref{fig:7} is apparently different from the semi-circle distribution.
We make four observations (see also Ref.~\cite{dgmc}):
(i) A $\delta$ peak at $\lambda_i=0$,
indicating the localized eigenstates that are produced by the dead-end vertices ;
(ii) $\delta$ peaks at $\lambda_i=\pm1$, indicating the existence of approximately infinite long chains;
(iii) A flat shape in the range $-1<\lambda_i<1$ except for $\lambda_i=0$,
indicating the existence of long chains constructed from weakly connected nodes, i.e.,
nodes with small degrees;
and (iv) A fat tail.

The log-log plot of the eigenvalue distribution
is shown in the right panel of Fig.~\ref{fig:7} (see also Ref.~\cite{sfa}).
In this figure the horizontal axis is the absolute value of eigenvalues
$|\lambda_i|$, and the vertical axis is the cumulative probability.
The plus symbols represent the distribution of the negative
eigenvalues, and the cross symbols represent that of the positive eigenvalues.
This figure shows that the shape of distribution in the positive eigenvalue range
and that in the negative eigenvalue range are almost the same.
The linear fitting with the least square method to the tail part
is shown by the solid line in the figure.
This fitting suggests that the PDF of the eigenvalue
$\rho(\lambda)$ is asymptotically given by
\[
\rho(\lambda)\propto\left|\lambda\right|^{-\delta},
\]
where the exponent $\delta=2.6$.
If we compare the values of $\gamma$ and $\delta$, we can find
the scaling law: $\delta=2\gamma-1$.

It has recently been shown under an effective medium approximation that
the PDF of eigenvalue $\rho(\lambda)$ is asymptotically represented by
that of the degree distribution $p(k)$:
\[
\rho(\lambda)\simeq 2\left|\lambda\right| p(\lambda^2),
\]
if the network has a local tree-like structure \cite{dgmc}.
Therefore, if $p(k)$ asymptotically follows the power law distribution,
$\rho(\lambda)$ also asymptotically follows the power law distribution,
and we can obtain the scaling relation $\delta=2\gamma-1$.
In addition, the local tree-like structure is guaranteed by the right panel of Fig.~\ref{fig:6}.

\section{Correlation between degree and company's total assets and age}
\label{sec:4}
It is interesting to construct models that can explain the topology of shareholding
networks.
However, in this section, we consider the correlation between the degree and 
the company's total assets and age.
In many complex networks, it is difficult to quantitatively characterize the nature of nodes.
However, in the case of economic networks, especially networks constructed of companies, we can
obtain the nature of nodes quantitatively.
We consider this to be a remarkable characteristic of business networks,
and this allows us to understand networks in terms of the company's growth.
This is the reason why we consider the correlation between the degree and the company's total
assets and age.

\begin{figure}[!t]
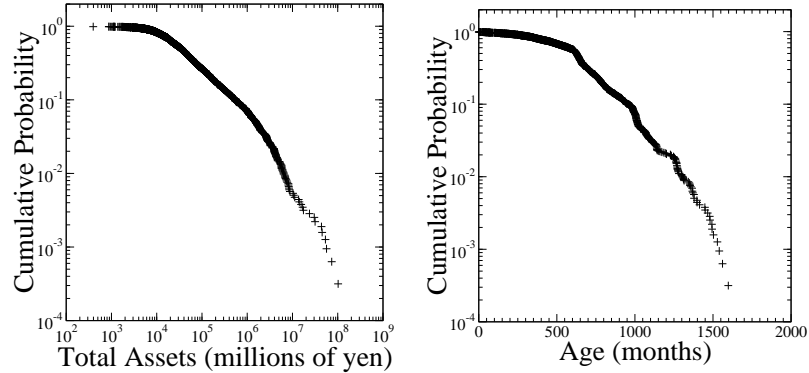

\centerline{
\begin{minipage}{0.43\linewidth}
\includegraphics[width=\linewidth]{rank_asset.eps}
\end{minipage}
\hspace{2mm}
\begin{minipage}{0.43\linewidth}
\includegraphics[width=\linewidth]{rank_age.eps}
\end{minipage}
}
\caption{A log-log plot of the distribution of the company's total assets (left) and
a semi-log plot of the company's age (right).}
\label{fig:8}
\end{figure}

The log-log plot of the distribution of the company's asset is shown in the left panel of Fig.~\ref{fig:8}.
In this figure, the horizontal axis is the assets with the unit of millions of yen, and the vertical axis is the
cumulative probability. This figure shows that the distribution in the intermediate range
follows the power law distribution. In this case, the distribution is for companies listed on the stock
market or the over-the-counter market. The completeness of data is a problem in extracting
the true nature of the total asset distribution. However, it has been shown that data satisfying this completeness shows more clearly the power law distribution of the total assets \cite{fgags}. 

The semi-log plot of the distribution of the company's age is shown in the right panel of Fig.~\ref{fig:8}.
In this figure, the horizontal axis is the age with the unit of months, and the vertical axis is the
cumulative probability. This figure shows that the distribution follows approximately
the exponential distribution.
It is expected that the age of companies has a relation with their lifetime, and
it is also clarified that the lifetime of bankrupted companies follows the exponential
distribution \cite{fujiwara}.

\begin{figure}[!t]
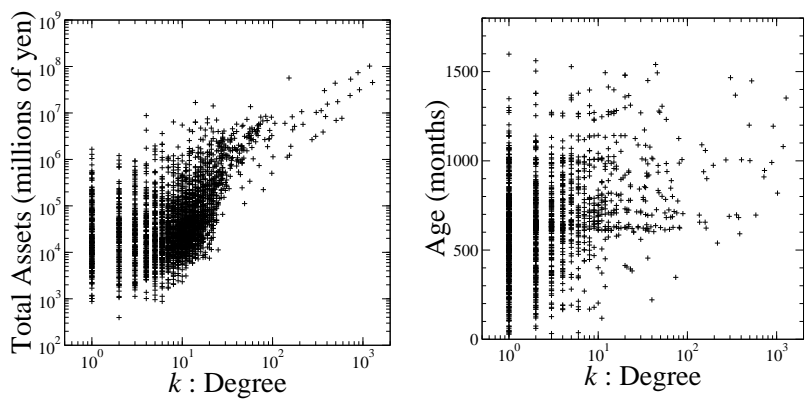

\centerline{
\begin{minipage}{0.43\linewidth}
\includegraphics[width=\linewidth]{deg_asset.eps}
\end{minipage}
\hspace{2mm}
\begin{minipage}{0.43\linewidth}
\includegraphics[width=\linewidth]{out_ltime.eps}
\end{minipage}
}
\caption{A log-log plot of the correlation between the degree and the company's total assets (left) and
a semi-log plot of the correlation between the degree and the company's age (right).}
\label{fig:9}
\end{figure}

The log-log plot of the correlation between degrees and total assets is shown in the left panel
of Fig.~\ref{fig:9}.
In this figure the horizontal axis is the degree, and the vertical axis is
the company's total assets with the unit of millions of yen. This figure shows that the degree and the total assets
positively correlate.

The semi-log plot of the correlation between degrees and the company's age is shown
in the right panel of Fig.~\ref{fig:9}.
In this figure the horizontal axis is the degree, and the vertical axis is the company's
age with the unit of months. This figure shows that the degree and the company's age
have no correlation.

These two results suggest that the degree of companies has a relation with their total assets,
but no relation with their age. This result means that the size of the company
is an important factor to consider with regard to growing economic networks, but
the age of the company is not.
Old companies are not necessarily big companies.
Hence knowing the dynamics of the company's growth is a key concept in considering
growing economic networks \cite{asf}.

\section{Summary}
\label{sec:5}
In this paper we considered the Japanese shareholding network at the end of March 2002,
and found some of the characteristics of this network.
However, there are many unknown facts about the characteristics of shareholding networks.
For example, these include time series changes of networks, the aspect of weighted networks,
flows in networks, and the centrality of networks.
Together with these studies, it is also important to study the dynamics of a
company's growth. It is expected that the dynamics of economic networks cam be explained
in terms of the dynamics of the company's growth.

\section*{Acknowledgements}
Wataru Souma and Yoshi Fujiwara are supported in part by the National Institute of Information and Communications
Technology.
We are also supported in part by a Grant-in-Aid for Scientific
Research (\#15201038) from the Ministry of Education, Culture,
Sports, Science and Technology.

\printindex
\end{document}